\documentclass[conference]{IEEEtran}
\ifCLASSINFOpdf
\else
\fi
\usepackage{url}
\usepackage{amsmath}
\usepackage{algpseudocode}
\usepackage{algorithm}
\usepackage{hhline}
\usepackage{graphicx}
\graphicspath{ {figures/} }
\usepackage{svg}
 \usepackage{booktabs}
 \usepackage{balance}
 \usepackage{url}
 \usepackage{hyperref}
 
 \usepackage[utf8]{inputenc}
 \usepackage{graphicx}
 \usepackage{makecell}
 \usepackage{pgfplots}
 
\begin{document}
%
\title{\textbf{LAN}TENNA: Exfiltrating Data from Air-Gapped Networks via Ethernet Cables}

\author{\IEEEauthorblockN{Mordechai Guri}
\IEEEauthorblockA{Ben-Gurion University of the Negev, Israel\\Cyber-Security Research Center\\
gurim@post.bgu.ac.il\\ air-gap research page:  \url{http://www.covertchannels.com} }}
 


%


\maketitle

\begin{abstract}
Air-gapped networks are wired with Ethernet cables since wireless connections are strictly prohibited.

In this paper we present LANTENNA - a new type of electromagnetic attack allowing adversaries to leak sensitive data from isolated, air-gapped networks.
Malicious code in air-gapped computers gathers sensitive data and then encodes it over radio waves emanating from the Ethernet cables, using them as antennas. A nearby receiving device can intercept the signals wirelessly, decode the data, and send it to the attacker. We discuss the exfiltration techniques, examine the covert channel characteristics, and provide implementation details. Notably, the malicious code can run in an ordinary user-mode process and successfully operate from within a virtual machine. We evaluate the covert channel in different scenarios and present a set of countermeasures. Our experiments show that with the LANTENNA attack, data can be exfiltrated from air-gapped computers to a distance of several meters away.

\end{abstract}


\renewcommand\IEEEkeywordsname{keywords}

\begin{IEEEkeywords}
	air-gap, exfiltration, covert channels, data leakage, Ethernet, LAN, electromagnetic.
\end{IEEEkeywords}

%
\IEEEpeerreviewmaketitle

\section{Introduction}
Information is the most valuable asset of modern organizations. Accordingly, adversaries spend a lot of resources and efforts to put their hands on the target information, usually documents and databases. After reaching the data, the attackers exfiltrate the information outside the boundaries of the organization. This is usually done in the form of covert communication channels within Internet protocols such as HTTPS, FTP, SMTP, and so on. Many massive data leakage incidents have been reported in the last decade. For example, In 2020, Microsoft disclosed a data breach event that occurred due to misconfigured security rules. According to the reports, 250 million records with personal information such as emails, IP addresses, and other details, were exposed \cite{AccessMi9:online}. In August 2021, Microsoft warned thousands of Azure users that their data might have been exposed through a database vulnerability named ChaosDB \cite{ChaosDBU63:online}. According to reports from the Identity Theft Resource Center (ITRC) and the U.S. Department of Health and Human Services, more than 98 million people were impacted by data breaches in the first half of 2021 \cite{Protecti60:online}.
 
\subsection{Air-Gap Networks}
Due to the increased risk of information leakage, when sensitive data is involved, an organization may move to so-called \textit{air-gap} isolation. Air-gapped computers are wholly separated from external wide area networks (WAN) such as the Internet \cite{May2018G94:online}. Many modern industries maintain their data within air-gapped networks, including financial, defense, and critical infrastructure sectors. Classified networks of military contractors and intelligence agencies may have air-gapped networks in place. E.g., the SIPRNet (Secret Internet Protocol Router Network) is a system of isolated and interconnected networks used by the U.S. Department of Defense to exchange classified information \cite{Storefro90:online}.

\subsection{Air-Gap Penetration} 
While theoretically air-gapped networks provide ultimate protection from cyber threats, in practice, it has been proven that even air-gapped networks are not immune to attacks. In 2008, the Agent.BTZ worm compromised classified and unclassified networks in the United States Central Command \cite{AgentBTZ65:online}. Another example is Stuxnet, the virus that infected the enriching uranium in Natanz nuclear facility. In this case, the malware was reportedly delivered via a thumb drive. Motivated adversaries can use sophisticated attacks to breach highly secure networks, such as compromising elements in the supply chain and infecting third-party software. Another attack tactic is to use malicious insiders in a so-called 'evil maid attack,' or exploit deceived insiders within the organization \cite{case2016analysis}\cite{PostDrea48:online}. 

In 2019, the Kudankulam Nuclear Power Plant in India was the target of a successful cyberattack earlier that year  \cite{AnIndian12:online}. In December 2020 SolarWinds breach, the hackers gained access to thousands of companies and government offices that used its products. The incident that was referred to as ``the largest and most sophisticated attack the world has ever seen" involved a highly evasive backdoor implanted within the company products. These types of techniques allow attackers to insert targeted malware into highly secured networks and environments. 

\subsection{Air-Gap Covert Channels}
After the attackers implanted their advanced persistent threat (APT) in the target network, they move on to the next phases of the attack kill chain. Initially, sensitive information is gathered: documents, images, keylogging, encryption keys, or databases. If the network is connected to the Internet (directly or via a virtual private network (VPN)), the data is exfiltrated through covert channels within known Internet protocols (e.g., HTTPS, FTP, SSH, and SMTP \cite{zander2007survey}). However, if the network is air-gapped, the attackers must exploit nonstandard communication techniques to exfiltrate the data outward, methods which are also referred to as air-gap covert channels \cite{Guri:2018:BAM:3200906.3177230}. Adversaries can exploit radio waves emanated from internal electronic components to transmit data \cite{guri2014airhopper,kuhn1998soft,kuhn2002compromising,vuagnoux2009compromising,guri2015gsmem}. They may also use the status LEDs on desktop computers to covertly transmit information  \cite{loughry2002information,Guri2017,Guri2017a}. Acoustic \cite{carrara2014acoustic,guri2020fansmitter}, thermal \cite{Guri2015a}, magnetic \cite{GURI2021115}, electric \cite{guri2019powerhammer}, and seismic air-gap covert channels have also been introduced over the years \cite{Carrara2016}.

\subsection{Our Contribution}
This paper introduces LANTENNA - a new type of electromagnetic attack that exploits the Ethernet networking cables to leak data wirelessly from air-gapped networks. Malware executed in a compromised workstation or server can regulate the electromagnetic waves emanated from an Ethernet cable, effectively use it as a transmitting antenna.
We show that any type of binary data can be modulated on top of the generated radio signals. We also show that a standard software-defined radio (SDR) receiver in the area can decode the information and then deliver it to the attacker via the Internet.

The following sections are organized as follows: Related work is discussed in Section \ref{sec:related}. The adversarial attack model is introduced in Section \ref{sec:attack}. Technical background is provided in Section \ref{sec:tech}. Sections \ref{sec:trans} and \ref{sec:rec}, respectively, describe the signal generation, and data transmission and reception. In Section \ref{sec:eval} we present the evaluation results. We discuss the possible countermeasures in Section \ref{sec:counter}, and we conclude in Section \ref{sec:conclusion}.

\section{Related Work}
\label{sec:related}
Certain malware such as the Conficker worm could move between computers via USB thumb drives \cite{shin2010conficker}. Kuhn showed that it is possible to exploit the electromagnetic emissions from the computer display unit to conceal data \cite{kuhn1998soft}. AirHopper, presented in 2014, is a malware capable of leaking data from air-gapped computers to a nearby smartphone via FM radio waves emitted from the screen cable \cite{guri2014airhopper,guri2017bridging}. In 2015, Guri et al. presented GSMem \cite{guri2015gsmem}, malware that transmits data from air-gapped computers to nearby mobile phones using cellular frequencies. USBee is malware that uses the USB data buses to generate electromagnetic signals \cite{guri2016usbee}. To prevent electromagnetic leakage, Faraday cages can be used to shield sensitive systems. Guri et al. presented ODINI \cite{guri2019odini} and MAGNETO \cite{GURI2021115}, two types of malware that can exfiltrate data from Faraday-caged air-gapped computers via magnetic fields generated by the computer's CPU. With MAGNETO, the authors used the magnetic sensor integrated into smartphones to receive covert signals. In 2019, researchers showed how to leak data from air-gapped computers by modulating binary information on the power lines \cite{guri2019powerhammer}. 

Several studies have proposed the use of optical emanations from computers for covert communication. Loughry and Guri demonstrated  the use of keyboard LEDs \cite{loughry2002information}\cite{guri2019ctrl}. Guri used the hard drive indicator LED \cite{Guri2017}, router and switch LEDs \cite{guri2018xled}, and security cameras and their IR LEDs \cite{guri2019air}, in order to exfiltrate data from air-gapped networks.

BitWhisper \cite{guri2015bitwhisper} is a thermal-based covert channel enabling bidirectional communication between air-gapped computers by hiding data in temperature changes.

Hanspach \cite{hanspach2014covert} used inaudible sound to establish a covert channel between air-gapped laptops equipped with speakers and microphones. Guri et al. introduced Fansmitter \cite{guri2020fansmitter}, Diskfiltration \cite{guri2017acoustic}, and CD-LEAK \cite{guri2020cd} malware which facilitates the exfiltration of data from an air-gapped computer via noise intentionally generated from the PC fans, hard disk drives, and CD/DVD drives. 

Another type of attack on the air-gapped system is the side channels. In this form of attack, adversaries could extract various information from systems remotely, via electromagnetic \cite{yilmaz2019electromagnetic}, acoustic \cite {genkin2014rsa}, and optical \cite{nassiglowworm} information leakage. 

\section{Attack Model}
\label{sec:attack}
The adversarial attack model consists of two main steps: infecting the air-gapped environment and data exfiltration. In the first step of the attack, the air-gapped environment is infected with malware, usually in the form of APT.

\subsection{Reconnaissance and Infection}
The APT Kill chain model was developed by Lockheed Martin that categorizes seven stages of targeted cyber attacks. The seven common phases of APT intrusions are reconnaissance, weaponization, delivery, exploitation, installation, Command \& Control, and data exfiltration. In the context of our work, the relevant phases to discuss are reconnaissance, delivery, and exfiltration.

In a reconnaissance phase, the attackers collect as much information as possible on their target, using various tools and techniques \cite{bahrami2019cyber}. After defining the initial target, attackers might install malware on the network via different infection vectors: supply chain attacks, contaminated USB drives, social engineering techniques, stolen credentials, or by using malicious insiders or deceived employees. 

Note that an infection of highly secure networks is proven to be feasible, as demonstrated by many incidents in the last decade \cite {langner2011stuxnet}, \cite{grant2009cyber},  \cite{TheEpicT20:online,RedOcto50:online,AFannyEq68:online}. At that point, the APT goal is to escalate privileges and spread in the network to strengthen its foothold in the organization.

\subsection{Data Exfiltration}
As a part of the APT exfiltration phase, the attacker might gather data from the compromised computers. The stolen information can be documents, databases, access credentials, encryption keys, and so on. 

\subsubsection{Data transmission}
Once the data is collected, the malware exfiltrates it using the covert channel. In the case of LANTENNA attack, it modulates the data and transmits wirelessly via the radio waves emanated from the Ethernet cables. 

\subsubsection{Data reception}
A nearby radio receiver can receive the covert transmission where it is decoded and send to an attacker. The receiving hardware can be carried by a malicious insider or hidden in the area.  

\begin{figure}  
	\centering
	\includegraphics[width=1\linewidth]{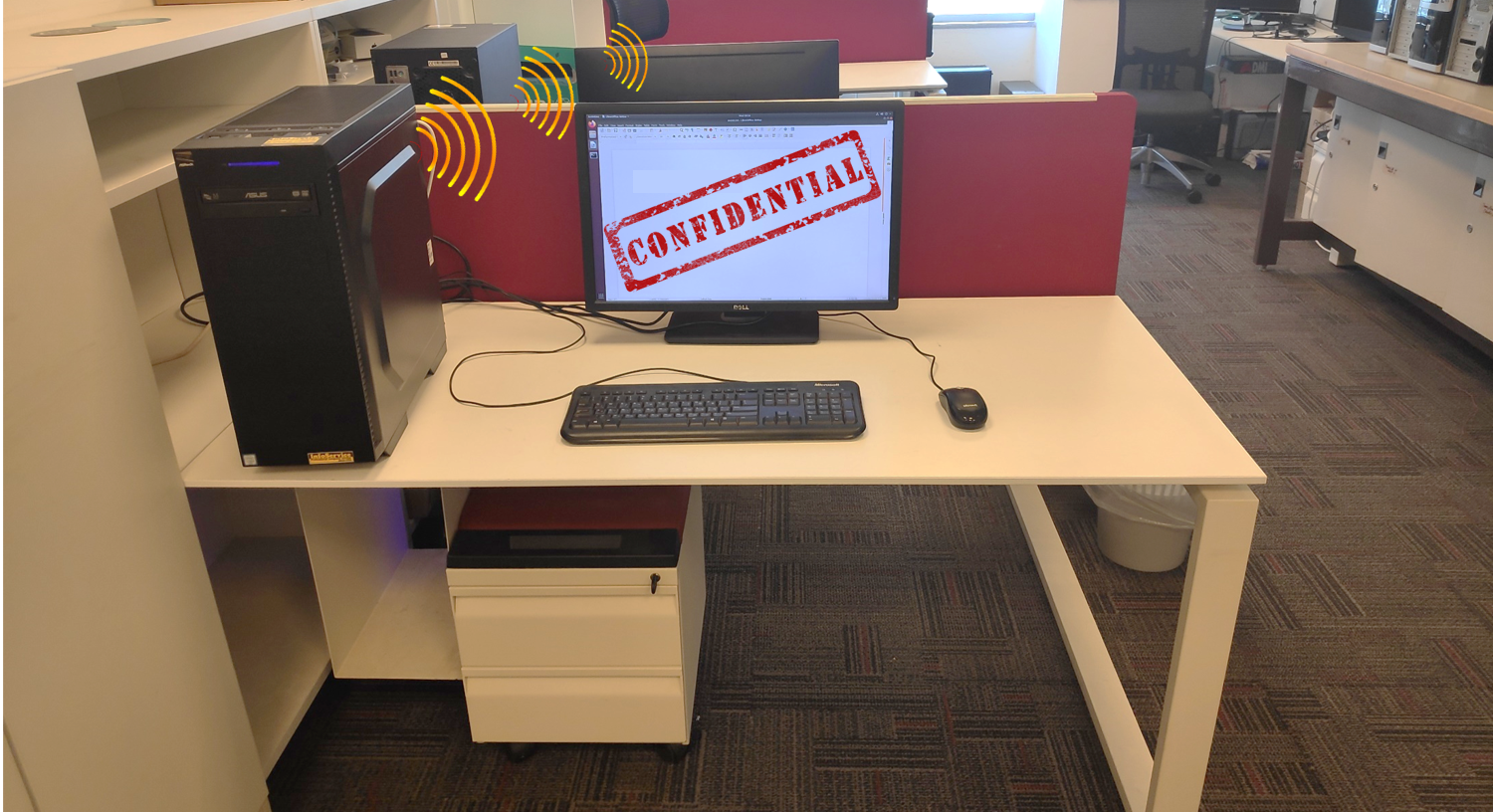}
	\caption{Illustration of the LANTENNA attack. Malware in the air-gapped network exploits the Ethernet cable, using it as an antenna to transmit radio signals. Binary information is modulated on top of the signals and intercepted by a nearby radio receiver.}
	\label{fig:attackmodel}
\end{figure}

The attack is illustrated in Figure \ref{fig:attackmodel}.
Malware in the air-gapped workstation generates electromagnetic emissions from the Ethernet cable. Binary information is modulated on top of the signals and intercepted by a nearby radio receiver.

%

\section{Technical Background}
\label {sec:tech}
Ethernet cables connect networked devices such as workstations and servers, printers, cameras, routers, and switches within a local area network (LAN). The network cable consists of eight wires which are twisted into four pairs. Several categories (cat) of Ethernet cables define parameters such as working frequencies, shielding, and bandwidth. The commonly used network cables are Cat 5, Cat 5e, Cat 6, Cat 6a, Cat 7, and Cat 7a. The main parameters of each category are specified in Table \ref{tab:cablelist}.

\begin{itemize}
	\item {\textbf{Cat 5, Cat 5e}}. Cat 5 and Cat 5e (Category 5 enhanced), are similar at the physical level, and they are both working at a maximal frequency of 100 MHz. However, while Cat 5 cable supports network bandwidth of 10-100 Mbps, a Cat 5e cable supports networks bandwidth up to 1 Gbps. In addition, the wires in Cat 5e cables are twisted more tightly than those in the Cat5 cable, which makes them more protected to unwanted signal inference between communication channels (crosstalk).
	Cat 5e is currently the most commonly used cable for home and small office facilities.
	
	\item \textbf{{Cat 6}}. Cat 6 cables support networks bandwidth up to 1 Gbps. They are better shielded, which helps in the prevention of crosstalk and electromagnetic interference. Cat 6 cables support bandwidth up to 1 Gbps for a distance of 55 meters. Cat 6 is mostly used in Enterprise IT networking environments.
	
	\item {\textbf{Cat 6a}}. Cat 6a (category 6 augmented) cables support network bandwidth up to 10 Gbps. Cat 6a cables are shielded and could eliminate crosstalk and interference. Cat 6a is mostly used in Enterprise IT networking environments. 
	
	\item \textbf{{Cat 7, Cat 7a}}. Cat 7 and Cat 7a (Category 7 enhanced) cables support higher bandwidths up to 10-40 Gbps to a range of 15 meters. Cat 7 cables are shielded and contain four individually shielded pairs inside an overall shield. They are manufactured with a GG45 connector which is also compatible with the RJ45 Ethernet ports. Cat 7 and Cat 7a are mostly used in data centers and high-bandwidth networking facilities. 
\end{itemize}

Besides these Ethernet cables, there are less common types. Cat 3 is 10 Mbps unshielded cable with a maximum bandwidth of 16 MHz and is obsolete today. The new fast Cat 8 cable is an emerging technology. Cat 8 currently offers one of the highest performance Ethernet capabilities. It supports a bandwidth of 40 Gbps, and it is highly shielded. Cat 8 is mostly used in big data centers and high-bandwidth networking facilities. 

\begin{table*}[]
	\centering
	\caption{Ethernet cable categories}
	\label{tab:cablelist}
	\begin{tabular}{@{}llclllll@{}}
		\toprule
		\# & Cable & \multicolumn{1}{l}{Frequency} & Ethernet Signal                      & Shielding       & Connector  & Pairs & Usage              \\ \midrule
		1  & Cat 5 & 100 MHz                       & \multicolumn{1}{c}{10/100 Base T}    & Optional        & 8p8c, RJ45 & 4     & Home, small office \\
		2  & Cat 5e & 100 MHz                       & 10/100 Base T, 1 Gigabit Ethernet    & Optional        & 8p8c, RJ45 & 4     & Home, small office \\
		3  & Cat 6  & 250 MHz                       & 10/100 Base T, 1 Gigabit Ethernet    & Optional        & 8p8c, RJ45 & 4     & Enterprise IT      \\
		4  & Cat 6a & 500 MHz                       & 10/100 Base T, 1/10 Gigabit Ethernet & Optional        & 8p8c, RJ45 & 4     & Enterprise IT      \\
		5  & Cat 7  & 600 MHz                       & 10/100 Base T, 1/10 Gigabit Ethernet & Pairs + overall & GG45, TERA & 4     & Datacenters        \\
		6  & Cat 7a & 1000 MHz                      & 10/100 Base T, 1/10 Gigabit Ethernet & Pairs + overall & GG45, TERA & 4     & Datacenters        \\ \bottomrule
	\end{tabular}
\end{table*}

\section{Transmission}
\label{sec:trans}
In this section we present the signal generation techniques, data modulation, and data transmission protocol. 

\subsection{Signal Generation}
Ethernet cables are electromagnetic in various frequency bands and amplitudes. Note that this phenomenon was studied in previous work in the general context \cite{eroglu1999practical, chen2008analysis}. Other work exploited the unintentional emission from the Ethernet cables to extract side-channel information \cite{carbino2014side, carbino2015exploitation}. In our work, we \textit{intentionally} exploit and trigger the electromagnetic emanation for data modulation and covert communication. 

We used two techniques to regulate the electromagnetic signals emanated from the Ethernet cables: (1) Ethernet speed toggling and (2) raw packets transmission. 
%

\subsection{Ethernet Speed Toggling}
\begin{figure}  
	\centering
	\includegraphics[width=0.8\linewidth]{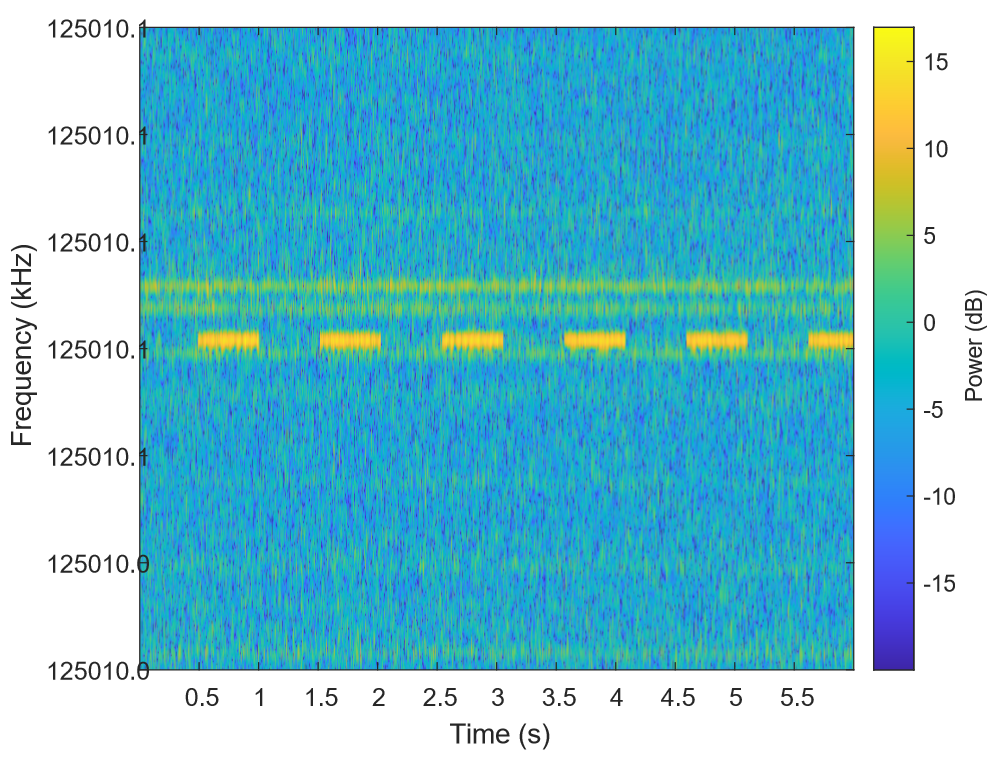}
	\includegraphics[width=0.8\linewidth]{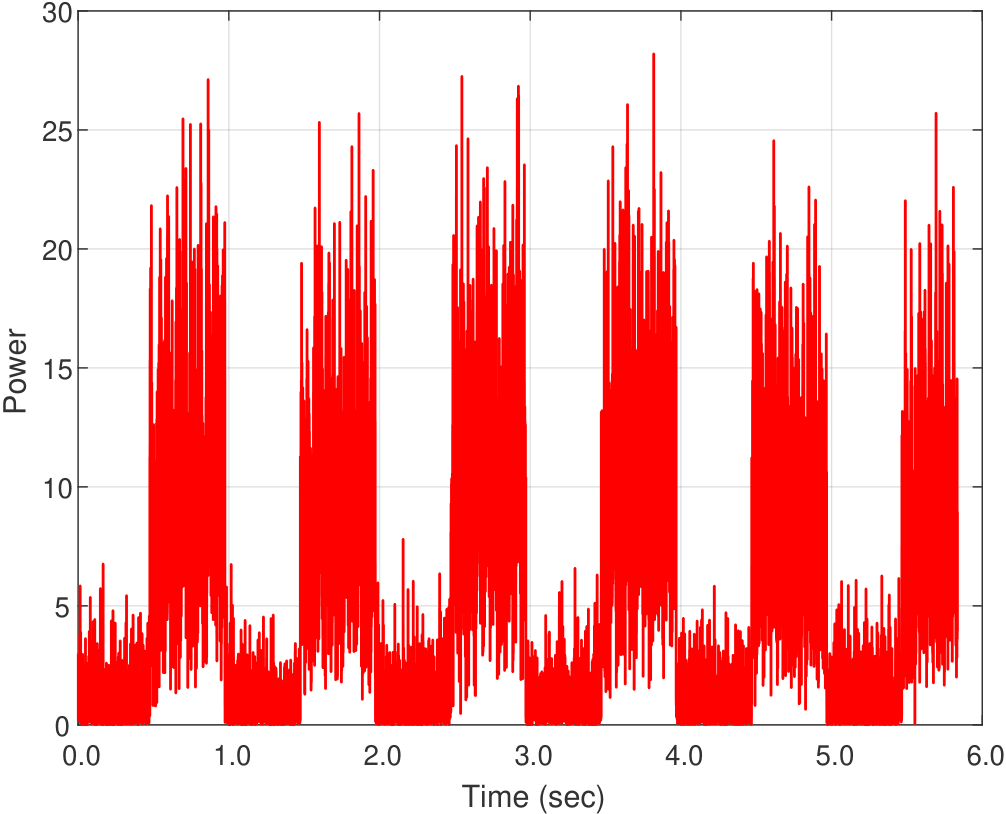}
	\caption{The waveform and spectrogram generated by a transmission of the alternating sequence `0101010...' from the Ethernet cable, using the Ethernet speed toggling.}
	\label{fig:trans1a}
\end{figure}

Ethernet cables emit electromagnetic waves in the frequency bands of 125 MHz and its harmonics (e.g., 250 MHz and 375 MHz). Note that the radiated frequency bands are determined by the operational frequency of the cable (e.g., 0-250 MHz for Cat 6, 0-500 MHz for Cat 6a, and 0-700 MHz for Cat 7 cables). We observed that changing the adapter speed or turning it on and off makes it possible to regulate the electromagnetic radiation and its amplitude. Figure \ref{fig:trans1a} shows the waveform and spectrogram generated by a transmission of the alternating sequence `10101010...' using the Ethernet speed toggling method with Cat 6 cable. In this case, the data was transmitted from an air-gapped computer through its Ethernet cable and received at a distance of 200 cm apart. As can be seen, the signal is wrapped around 125.010 MHz.

\subsection{Raw Packets Transmission}
\begin{figure}  
	\centering
	\includegraphics[width=0.8\linewidth]{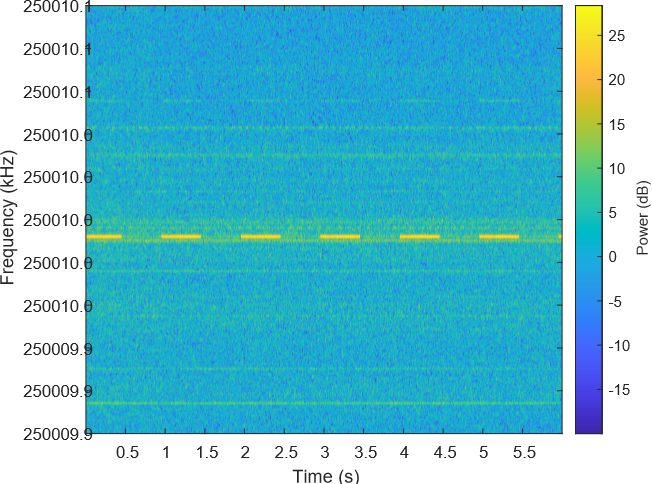}
	\includegraphics[width=0.8\linewidth]{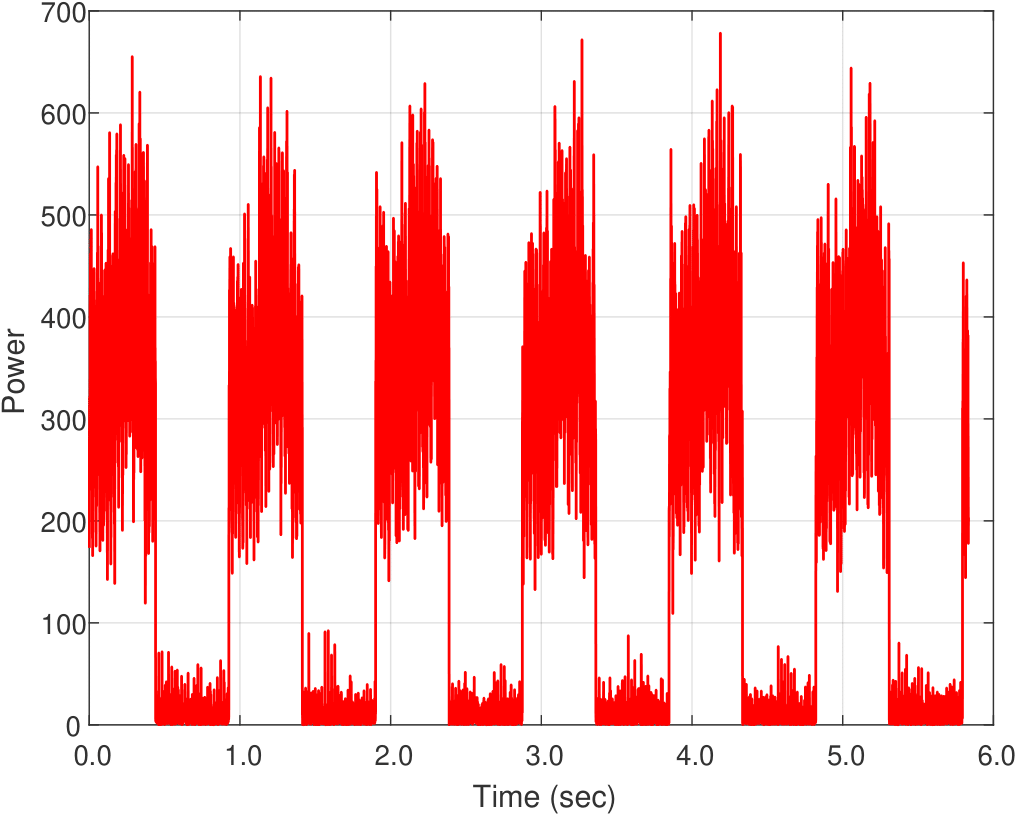}
	\caption{The waveform and spectrogram generated by a transmission of the alternating sequence `10101010...' from the Ethernet cable, using the raw packets transmission method.}
	\label{fig:trans2a}
\end{figure}

The data and link layers activities determine the current flow on the copper wires in the Ethernet cables. By sending raw UDP packets, we could trigger and regulate the emission from the Ethernet cable at its operational frequency. Figure \ref{fig:trans2a} shows the waveform and spectrogram generated by transmission of the alternating sequence `10101010...' using the raw packet transmission method. In this case, the data was transmitted from an air-gapped computer through its Ethernet cable and received at a distance of 200 cm apart. As can be seen, the signal is wrapped around 250.010 MHz and is narrower than the signal generated by the speed toggling method.

\begin{figure*}[t] 
	\centering
	\includegraphics[width=0.8\linewidth]{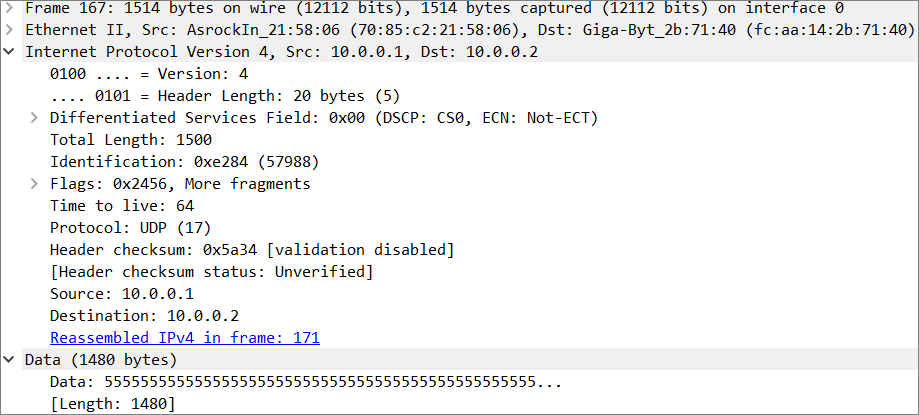}
	\caption{The UDP frame generates the electromagnetic emission, as shown in the Wireshark network analyzer.}
	\label{fig:frame}
\end{figure*}

The pseudo code of the modulator is shown in Algorithm 1. The \texttt{modulate} function receive the array of bits to transmit (\texttt{bits}) and the time of each bit (\texttt{bitTimeMillis}).
If the bit to transmit is '1', the function sends UDP packets to the network; otherwise, it sleeps for the same duration. Each UDP packet contains a payload of 1480 bytes. The payload consists of a sequence of the 'U' character, which is the alternate bits (\texttt{01010101}) in its binary representation. The full UDP frame, as shown in the Wireshark network protocol analyzer, is presented in Figure \ref{fig:frame}.
\begin{algorithm}
\caption{modulate(bits, bitTimeMillis)} 
\label{alg0} 
\begin{algorithmic}[1] 

    \State $ bitEndTime \gets getCurrentTimeMillis() $
    
    \For{$ bit\ in\ bits $}
        \State $ bitEndTime \gets bitEndTime + bitTimeMillis $
        \State $ halfBitEndTime \gets bitEndTime-bitTimeMillis/2 $
        \If {$ bit == 1 $} 
            \State $ sleep(bitTimeMillis/2) $ 
            \While{$getCurrentTimeMillis() <  bitEndTime$}
                \State $ sendPackets() $ 
            \EndWhile
        \Else
            \While{$getCurrentTimeMillis() <  halfBitEndTime $}
                \State $ sendPackets() $ 
            \EndWhile
            \State $ sleep(bitTimeMillis/2) $ 
        \EndIf
        
    \EndFor  
      
\end{algorithmic}
\end{algorithm}

\subsection{Encoding and Packets Frames}
\begin{figure}
	\centering
	\includegraphics[width=0.6\linewidth]{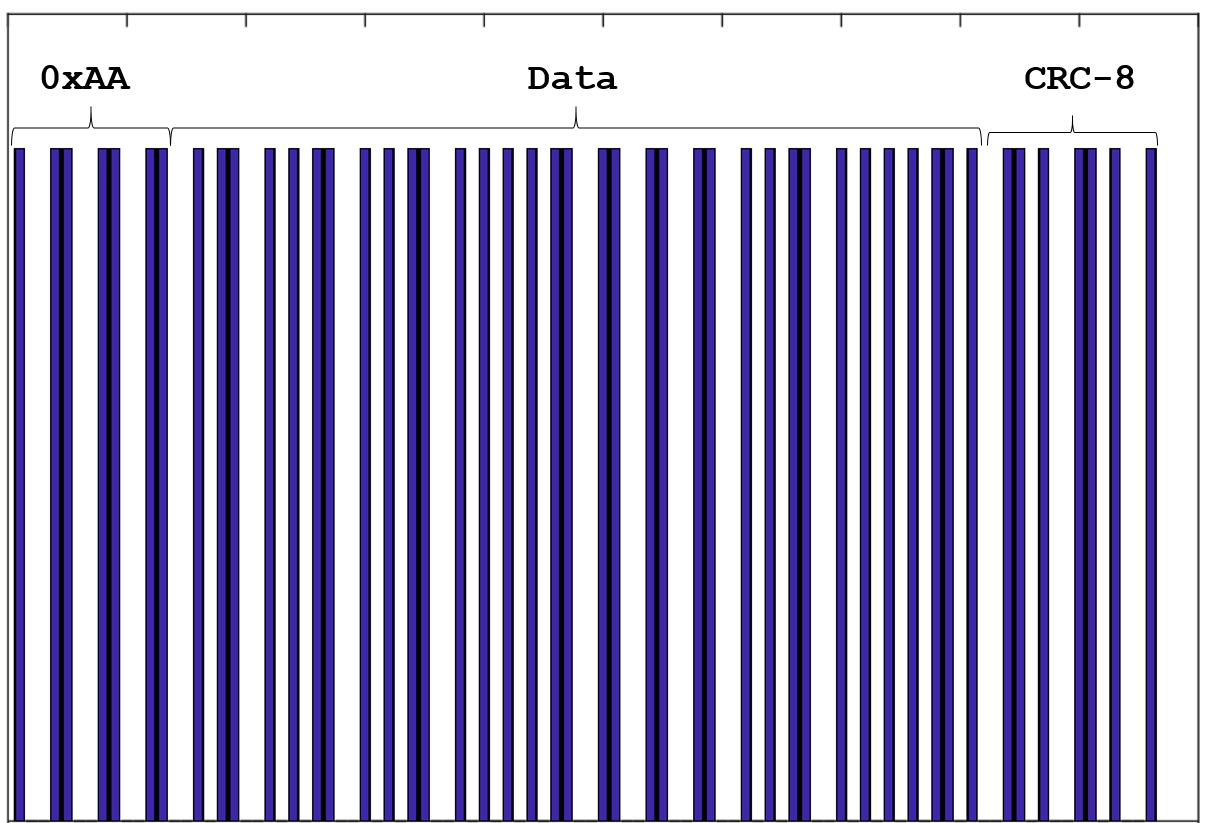}
	\caption{Transmission of a packet with enable (0xAA), data payload, and CRC-8 code.}
	\label{fig:message}
\end{figure}

Note that the amplitude of the signal may change over time, and hence, a simple OOK modulation fails during the reception. We used Manchester encoding since a bit is demodulated by analyzing the change in amplitude during both halves of the bit with no overall threshold is needed. Figure \ref{fig:message} depicts the Manchester encoding for the packet: enable = 0xAA, DATA = 'DATA', CRC8 = 0xB6.

\begin{itemize}
	\item {Enable.} The packet begins with a \texttt{0xAA} hex value. This sequence of \texttt{10101010} in binary allows the receiver to synchronize with the beginning of each packet and determine the carrier amplitude and one/zero thresholds. 
	
	\item {Data.} The payload is the raw binary data transmitted within the packet. It consists of 32 bits.  
	
	\item {CRC-8.} For error detection, we use the CRC-8 (a cyclic redundancy check) error detection algorithm. The CRC is calculated on the payload data and added at the end of each packet. If the received CRC and the calculated CRC differ, the packet is omitted on the receiver side. 
\end{itemize}

%
%
%

\section{Reception}
\label{sec:rec}

\subsection{Demodulation}
The pseudo-code of the demodulator is presented in Algorithm 2. We provide the implementation for software-defined radio (SDR) receivers.

\begin{algorithm}
\caption{demodulate(fileName, freq, sampleRate, bitTime, windowSize)} 
\label{alg0} 
\begin{algorithmic}[1] 
\State $ enabled \gets False $
\State $ windowsPerBit \gets bitTime*sampleRate/windowSize $
\State $ fileID = writeRtlSdrOutputToFile(fileName, \gets  freq, sampleRate) $
\\
\While { $ True $ }
    \State $ window = fileID.blockingRead(windowSize) $

    \State $ spectrum = fft(window)$
    \State $ sampleValue \gets getSignalAmplitude(spectrum) $
    \State $ samples.append(sampleValue) $

    \If {$not\ enabled$}
        \State $ enabled \gets detectEnable(samples) $
    \EndIf
    
    \While {$ enabled\ and\ enoughSamplesForBit( \gets samples, windowsPerBit) $}
        \State $ bit \gets samplesToBitManchester( \gets  samples, windowsPerBit) $
        \State $ output(bit) $
    \EndWhile
    
\EndWhile
      
\end{algorithmic}
\end{algorithm}

The \texttt{demodulate} function is based on sampling and processing the FFT information for the target frequency bands (125 MHz, 250 MHz, 375 MHz, and so on.). In lines 2-3, the SDR device is initialized, and the receiving buffer is configured with the frequency (in MHz) of the channel to monitor (\texttt{freq}), the sampling rate (\texttt{sampleRate}), and the buffer size (\texttt{windowsPerBit}). The demodulator samples the data in the target frequency and splits it into windows of \texttt{windowSize} size. The algorithm estimates the power spectral density for each window using Welch's method (lines 9-13). It then detects the \textit{enable sequence} (\texttt{10101010}) using the \texttt{detectEnable} routine. It then decodes the bit using Manchester scheme (\texttt{SampleToBitManchester}) and determines the thresholds (amplitudes) for '1' and '0' bits (lines 14-18). Finally, the bits are demodulated and added to the output vector (lines 18-21).    

\section{Evaluation}
\label{sec:eval}
In this section, we present the evaluation of the covert channel. We describe the experimental setup and test the different reception modes used to maintain the covert channel. 

\begin{table}[]
	\centering
	\caption{Ethernet cables used for the evaluation}
	\label{tab:cables}
	\begin{tabular}{@{}lll@{}}
		\toprule
		\#      & Color & Type        \\ \midrule
		cable-1 & White & CAT 5e UTP    \\
		cable-2 & Blue  & CAT 6A S/FTP \\
		cable-3 & Green & CAT 6A U/FTP \\ \bottomrule
	\end{tabular} 
\end{table}

\subsection{Experimental Setup}

\subsubsection{Receivers}
For the reception we used two types of software-defined radio (SDR) receivers, as specified in Table \ref{tab:Receivers}.
The R820T2 RTL-SDR is capable of sampling up to 16bit at narrow band and has RF coverage from 30 MHz to 1.8 GHz or more. The HackRF device has 1 MHz to 6 GHz operating frequency and 8-bit quadrature samples (8-bit I and 8-bit Q). Both receivers are compatible with GNU Radio, SDR\#, and others. We connected the receiver through the USB port to a laptop, with an Intel Core i7-4785T and Ubuntu 16.04.1 4.4.0 OS.

\begin{table*}[]
	\centering
	\caption{Receivers used in the evaluation}
	\label{tab:Receivers}
	\begin{tabular}{@{}lll@{}}
		\toprule
		Receiver \#   & Device          & Specs                                                                   \\ \midrule
		SDR-1 &  R820T2 RTL-SDR     & Frequency range from 30 MHz to 1.8 GHz                    \\
		SDR-2 &  HackRF & Frequency range from 1 MHz to 6 GHz, with software-controlled antenna port power (50 mA at 3.3 V) \\ \bottomrule
	\end{tabular}
\end{table*}

\subsubsection{Transmitters}
For the transmission, we used the three types of off-the-shelf workstations listed in Table \ref{tab:setup1}. The computers are equipped with 10/100/1000 Mbps Gigabit Ethernet card.
We tested three types of widely used Cat 5e and Cat 6A Ethernet cables listed in Table \ref{tab:cables1}. We also tested a laptop computer and an embedded device (Raspberry Pi) to evaluate the attack on these types of devices.


\begin{table*}[]
\centering
	\caption{The workstations used for the evaluation}
	\label{tab:setup1}
\begin{tabular}{||c | c | c||}
	
	\hline
	\thead{PC} & \thead{Hardware} & \thead{OS}  \\ [0.5ex]
	\hline\hline
	PC1 &
	\makecell{Lenovo ThinkCentre M93p, Intel Core i7-4785T, \\8GiB SODIMM DDR3 Synchronous 1600 MHz\\
		NIC: driver e1000e, device I217-LM (rev 04)} &
	\makecell{Ubuntu 16.04.1 \\4.4.0-modified} \\
	
	\hline
	PC2 &
	\makecell{ASRock X99 Extreme4, Intel Core i7-6900K, \\4 * 8GB PC4-1900 DDR4 SDRAM \\SK Hynix HMA81GU6AFR8N-UH\\
		NIC: driver e1000e, device I218-V (rev 05)} &
	\makecell{Ubuntu 18.04.2 5.0.0-25-generic} \\
	
	\hline
	PC3&
	\makecell{H97M-D3H, Intel Core i7-4790, \\4 * 4GB DIMM DDR3 1600MHz Hynix\\
		NIC: driver r8169, device RTL8111/8168/8411 (rev 06)} &
	\makecell{Ubuntu 18.04.1 \\4.15.0-72-generic}\\ 
		\hline
	Laptop &
	\makecell{Dell 0T6HHJ, Intel(R) Core(TM) i7-6600U CPU @ 2.60GHz		
			\\  8GB PC4-1900 DDR4 \\ 		
		NIC: driver e1000e, device I218-V} &
	\makecell{Ubuntu 18.04.3 LTS}\\ 	\hline
	
	Embedded &
	\makecell{ Raspberry Pi 3 Model, Model B} &
	\makecell{Raspberry Pi OS}\\ [1ex]

	\hline
\end{tabular}
\end{table*}

The following subsections present the results obtained for the two transmission methods.

\subsection{Ethernet Cables}
Table \ref{tab:cables1} shows the signal-to-noise ratio (SNR) levels of a transmission generated by toggling the Ethernet interface for different PC with different cables. As can be seen, the signal strength depends on the transmitting computer and the cable used. In this case, PC1 and PC3 with cable-1 and cable-2 yield the most vital signals.

\begin{table}[H]
	\centering
	\caption{The SNR of different cables}
	\label{tab:cables1}
	\begin{tabular}{||c | c | c | c||} 
		
		\hline
		\thead{} & \thead{cable-1} & \thead{cable-2} & \thead{cable-3}  \\ [0.5ex] 
		\hline\hline
		
		\hline
		\makecell{PC1} & 16.74 dB & 6.35 dB & 7.55 dB\\
		\hline
		\makecell{PC2} & 1.5 dB & 4.75 dB & 8.02 dB\\
		\hline
		\makecell{PC3} & 14.125 dB & 14 dB & 5.49 dB\\

		\hline
	\end{tabular}
\end{table} 

\subsection{Frequency bands}
Our experiments show that the exact frequency band is derived from the type of the transmitting device. Table \ref{tab:freq1} shows the frequency responses for the Ethernet speed toggling method for the PC, laptop, and embedded devices. The base frequency is wrapped around 250 MHz with differences of less than 0.1 MHz between them.


\begin{table}[]
	\centering
	\caption{Main frequency bands}
	\label{tab:freq1}
	\begin{tabular}{@{}llll@{}}
		\toprule
		& PC          & Laptop        & Embedded  \\ \midrule
		Frequency & 250.000 MHz & 249.99488 MHz & 250.00285 MHz\\ \bottomrule
	\end{tabular}
\end{table}

\subsubsection{Harmonics}
\begin{figure}
	\centering
	\includegraphics[width=0.8\linewidth]{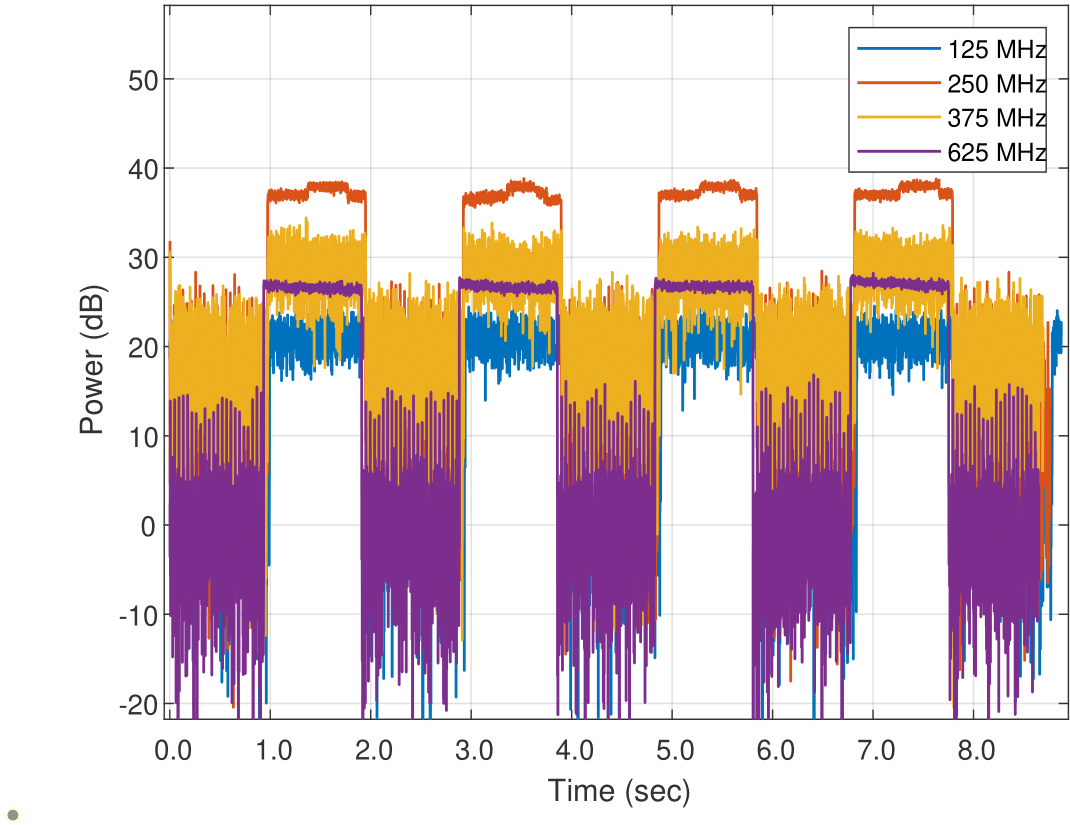}
	\caption{Harmonics of the signal generated from the PC1.}
	\label{fig:harminics}
\end{figure}

There are various harmonics for the main signal. Figure \ref{fig:harminics} shows the harmonics of the signal generated from  PC1. The 250 MHz is the strongest harmonics with 37 dB signal, while the 375 MHz, 125 MHz, and 625 MHz bands yield a weaker signal. In the context of the attack, it implies that it is optimal to calibrate the receiver device to the 250 MHz frequency band.

\subsection{Ethernet Speed Toggling}

\subsubsection{SNR}
Table \ref{tab:SNR1} shows the SNR values generated by the Ethernet speed toggling for the PC, laptop, and embedded transmitters. As can be seen, the PC and embedded signals were successfully received from a distance of 4m and 4.5m, respectively. The SNR values are decreased from 27 dB to 7 dB in PC and 12 dB to 3 dB in the embedded device. The laptop could generate a weak signal for a maximal distance of 50 cm with an SNR of 8 dB at most.      
\begin{table*}[]
	\centering
	\caption{SNR of the Ethernet speed toggling}
	\label{tab:SNR1}
	\begin{tabular}{@{}lllllllllll@{}}
		\toprule
		& 0 cm  & 50 cm  & 100 cm & 150 cm & 200 cm & 250 cm & 300 cm & 350 cm & 400 cm & 450 cm \\ \midrule
		PC        & 27 dB     & 15 dB     & 18 dB      & 14 dB      &  13 dB     & 7 dB      & 7 dB      & 6 dB      &   7 dB    & -      \\
		Laptop    & 8 dB     & 2.6 dB      & -      & -      & -      & -      & -      & -     & -      & -      \\
		Embedded & 12 dB & 6.5 dB & 6 dB   & 7 dB   & 5.5 dB & 5 dB   & 5 dB   & 4.5 dB & 3 dB   & 3 dB   \\ \bottomrule
	\end{tabular}
\end{table*}

\subsubsection{Speed}
Table \ref{tab:timing1} shows the transition response between link speeds of 10, 100, and 1000 for PC, laptop, and embedded devices. Note that for the Ethernet toggling method, the bit rate is derived directly from the transition time. As can be seen for the PC and laptop transmitters, an average of 4 seconds is required for a transition between different speeds and turning the interface on. However, shutting the interface down takes much less time, with 0.013-0.024 seconds on average. The embedded devices transition is much faster, with an average speed of 0.095-0.017 seconds for the 0-100 Mbps.    

\begin{table*}[]
	\centering
	\caption{Timing measurements of the Ethernet speed toggling}
	\label{tab:timing1}
	\begin{tabular}{@{}lccccc@{}}
		\toprule
		& \multicolumn{1}{l}{0-10 Mbps (up/down)} & \multicolumn{1}{l}{0-100 Mbps (up/down)} & \multicolumn{1}{l}{0-1000 Mbps (up/down)} & \multicolumn{1}{l}{10-100 Mbps (up/down)} & \multicolumn{1}{l}{100-1000 Mbps (up/down)} \\ \midrule
		PC        & 4 sec / 0.013 sec                       & 4 sec / 0.013 sec   sec                        & 4-6 sec / 0.013 sec                       & 4 sec / 4 sec                              & 4 sec / 4 sec                                \\
		Laptop    & 4 sec / 0.02 sec                           & 4 sec / 0.024 sec sec                        & 4 sec / 0.024 sec                           & 4 sec / 4 sec                              & 4 sec / 4 sec sec                               \\
		Embedded & 0.095 sec / 0.17 sec                                     & 0.095 sec / 0.17 sec                                         & -                                       & 0.081 sec / 0.072 sec                                          &  -                                          \\ \bottomrule
	\end{tabular}
\end{table*}

\subsection{Raw Packet Transmission}

\subsubsection{SNR}
\begin{figure}
	\centering
	\includegraphics[width=0.7\linewidth]{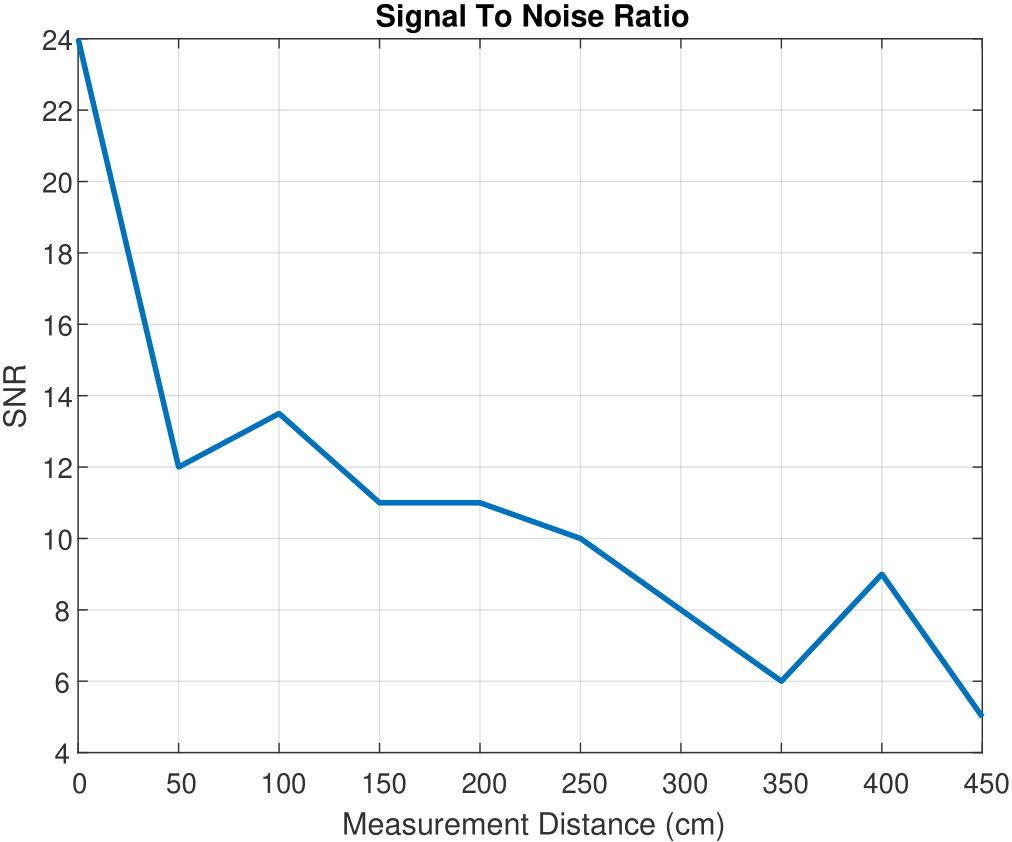}
	\caption{The SNR levels for the PC with the raw packet transmission method.}
	\label{fig:SNRPC}
\end{figure}

Table \ref{tab:SNR1} shows the SNR values generated by the raw packet transmission for the PC, laptop, and embedded transmitters. As can be seen, the signal for the PC was successfully received from a distance of 4.5m. As can be seen in Figure \ref {fig:SNRPC} the SNR values are decreased from 24 dB to 5 dB. The laptop and embedded devices could generate a weak signal for maximal distances of 1m and 1.5m, respectively, with SNR values of 4.77 dB and 8 dB at most.  

 \begin{table*}[]
 	\centering
 	\caption{SNR of the raw packet transmission}
 	\label{tab:SNR2}
 	\begin{tabular}{@{}lllllllllll@{}}
 		\toprule
 		& 0 cm  & 50 cm  & 100 cm & 150 cm & 200 cm & 250 cm & 300 cm & 350 cm & 400 cm & 450 cm \\ \midrule
 		PC        & 24 dB     & 12 dB     & 13.5 dB      & 11 dB      &  11 dB     & 10 dB      & 8 dB      & 6 dB      &   9 dB    & 5 dB      \\
 		Laptop    & 5.5 dB     & 5 dB     & 4.77 dB      & -      & -      & -      & -      & -      & -      & -      \\
 		Embedded & 20 dB & 11 dB  & 10 dB      & 8 dB      &  -      & -      & -     & -      & -      & -      \\
 		\bottomrule
 	\end{tabular}
 \end{table*}

\subsubsection{Speed}
Tables \ref{tab:BERPC}, \ref{tab:BERLPTOP} and \ref{tab:beriot} show the bit error rates (BER) for PC, laptop, and embedded devices, respectively, with the raw packets transmission method. for the PC, transmission rates of 1 bit/sec and 5 bit/sec maintained 0\% errors up to a distance of 3m. With a transmission rate of 10 bit/sec we maintained 12.5\% errors up to a distance of 3m.
For the embedded device, transmission rates of 1 bit/sec and 5 bit/sec maintained 0\% errors up to a distance of 3.5m. However, we maintained 0\% errors to a distance of 1m only for a transmission rate of 1 bit/sec with a laptop. With transmission rates of 5 bit/sec and 10 bit/sec, we could reach short 1m and 0.5m distances, respectively.

\begin{table*}
	\centering
	\caption{BER for the PC, with the raw packets transmission}
	\label{tab:BERPC}
	\begin{tabular}{@{}llllll@{}}
		\toprule
		& 0 cm            & 50 cm           & 100 cm          & 200 cm          & 300 cm          \\ \midrule
		1 bit/sec  & 0\% (no errors) & 0\% (no errors) & 0\% (no errors) & 0\% (no errors) & 0\% (no errors) \\
		5 bit/sec  & 0\% (no errors) & 0\% (no errors) & 0\% (no errors) & 0\% (no errors) & 0\% (no errors) \\
		10 bit/sec & 0\% (no errors) & 0\% (no errors) & 12.5\%          & 12.5\%          & 12.5\%          \\ \bottomrule
	\end{tabular}
\end{table*}

\begin{table}[]
	\centering
	\caption{BER for the laptop, with the raw packets transmission}
	\label{tab:BERLPTOP}
	\begin{tabular}{@{}llll@{}}
		\toprule
		& 0 cm            & 50 cm           & 100 cm          \\ \midrule
		1 bit/sec  & 0\% (no errors) & 0\% (no errors) & 0\% (no errors) \\
		5 bit/sec  & 0\% (no errors) & 0\% (no errors) & 12.5\%          \\
		10 bit/sec & 0\% (no errors) & 12.5\%          & 37.5\%          \\ \bottomrule
	\end{tabular}
\end{table}


\begin{table*}[]
	\centering
	\caption{BER for the embedded device, with the raw packets transmission}
	\label{tab:beriot}
	\begin{tabular}{lllllll}
		\hline
		& 0 cm            & 50 cm           & 100 cm          & 200 cm          & 300 cm          & 350 cm          \\ \hline
		5 bit/sec  & 0\% (no errors) & 0\% (no errors) & 0\% (no errors) & 0\% (no errors) & 0\% (no errors) & 0\% (no errors) \\
		10 bit/sec & 0\% (no errors) & 0\% (no errors) & 0\% (no errors) & 0\% (no errors) & 0\% (no errors) & 0\% (no errors) \\ \hline
	\end{tabular}
\end{table*}


%
%
%

\subsection{Virtual Machines (VMs)}
We examined whether the covert channel could be launched from within 
virtual machines. Since virtualization has become a standard in many IT environments today, the malicious code would likely run in guest OS. One of the properties of visualization technologies is the isolation of hardware resources. Hypervisors/virtual machine monitors (VMMs) provide a layer of abstraction between the virtual machine and the physical hardware, including the network interface card. The architecture of virtual machine networking uses the concept of virtual network adapters. A virtual network adapter is maintained by the hypervisor and exposed to the guest via kernel drivers. In the context of our covert channel, there is two common networking configurations for virtual machines:

\begin{itemize}
	\item {NAT mode}. In this mode, the guest OS on a VM communicates with other hosts in the local area network through a virtual NAT (Network Address Translation). Other workstations and networked devices can be accessed from a guest OS. However, the IP address of the VM is assigned via DHCP, and the external network is exposed only to the IP of the host machine. 
	
	\item {Bridge mode}. In this mode, the virtual network adapter is connected to the physical network adapter of the host machine. The network traffic is sent and received directly from/to the real network adapter without encapsulation, modification, or routing.   
\end{itemize}

\begin{figure}
 	\centering
 	\includegraphics[width=1\linewidth]{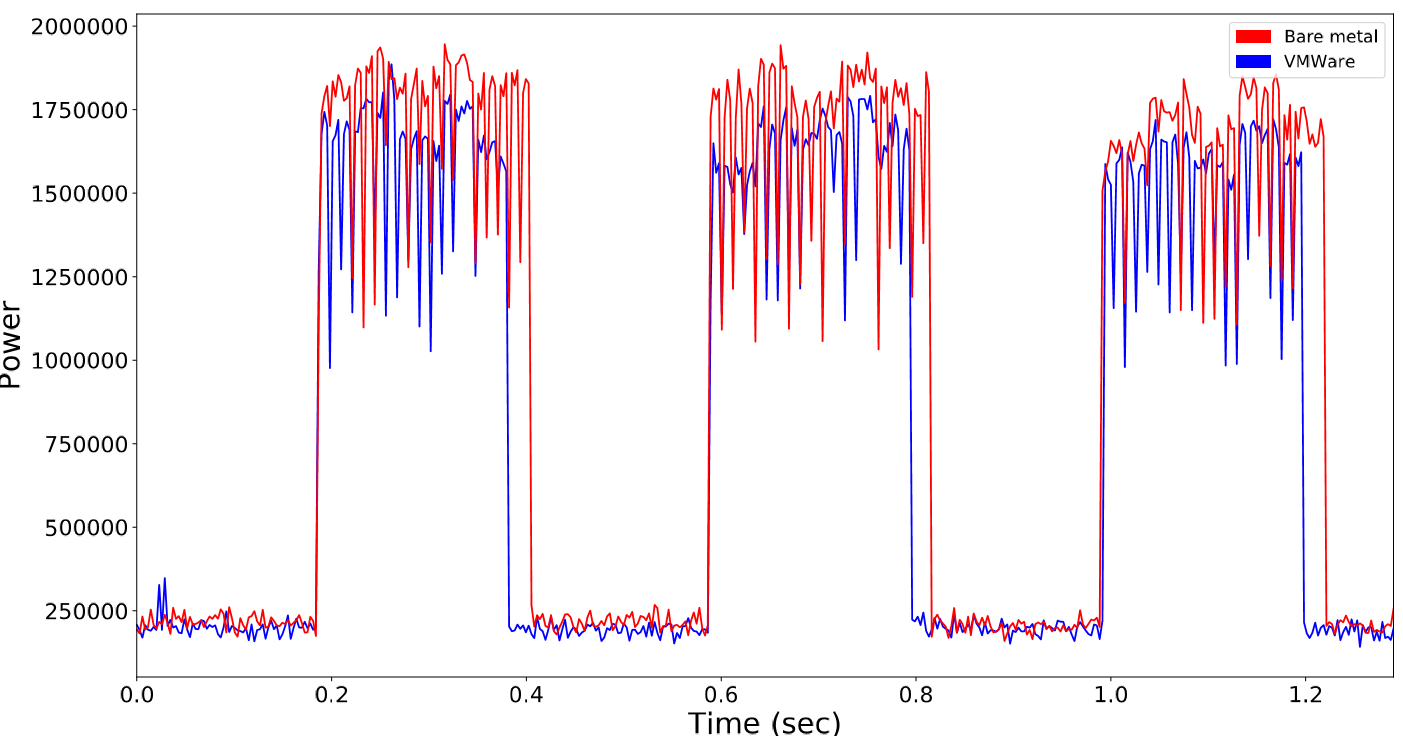}
 	\caption{Signal generated from a bare metal and a WMware virtual machine.}
 	\label{fig:VMM}
\end{figure}

A process executed in a VM can access the virtual networks cards assigned to the VM. In regard to the covert channel, it implies the malicious code can not disable or change the speed of the \textit{physical} network interface and hence, can not control the electromagnetic emission using the network toggling technique. However, since UDP packets can be delivered to the network, the raw packet transmission technique can still be used in both NAT and bridge modes.
We compared the covert channel using a bare-metal machine and VMware VMM in the NAT and bridge modes in PC3. Our experiments show that the covert signals can be maintained from within virtual machines. Figure \ref{fig:VMM} shows the electromagnetic signals generated using the raw packet transmission with bridged network configurations on bare metal and VMWare workstations. As can be seen, the execution on a bare metal yields a slightly stronger signal than a VM, mainly due to the delay in packet transmission caused by the hypervisor.  

\subsection{Evasion}
The Ethernet speed toggling method generates strong signals, but it is less evasive than the raw packet transmission. First, it required root privileges to perform the changes to the network interfaces speeds. Technically, it requires the process to run in root privileges or exploit a privilege escalation vulnerability. Both techniques can be monitored and detected by modern intrusion detection systems. The raw packet transmission method can be executed as an ordinary user-level process. Transmitting UDP packets doesn't require higher privileges or interfering with the OS routing table. In addition, it is possible to evade detection at the network level by sending the raw UDP traffic within other legitimate UDP traffic.

\section{Countermeasures}
\label{sec:counter}
There are several defensive measures that can be taken against the LANTENNA covert channel. 
 
\subsection{Separation}
The NATO telecommunication security standards (e.g., NSTISSAM TEMPEST/2-95 \cite{NSTISSAM75:online}) propose zone separation to protect against TEMPEST (Telecommunications Electronics Materials Protected from Emanating Spurious Transmissions) threats and other types of radiated energy attacks. In our case any radio receiver should be banned from the area of air-gapped networks.

\subsection{Detection} 
\begin{figure}  
	\centering
	\includegraphics[width=0.6\linewidth]{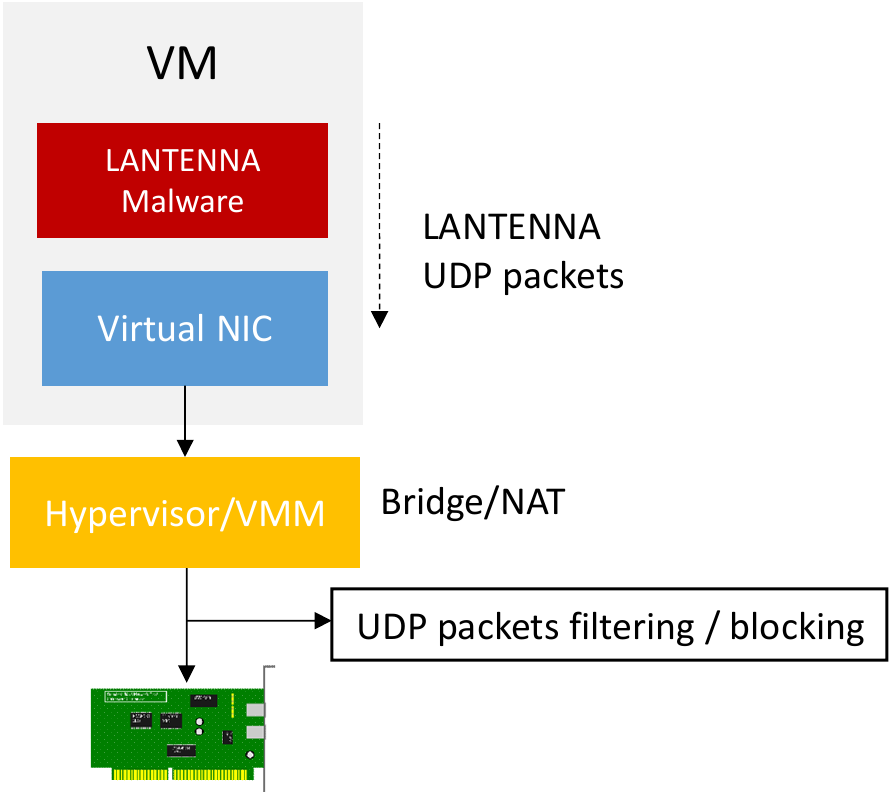}
	\caption{Hypervisor-level detection and prevention.}
	\label{fig:filter}
\end{figure}

For the Ethernet speed toggling method, it is possible to monitor the network interface card link activity at the user and kernel levels. In our case, any change of the link state should trigger an alert. E.g., usage of \texttt{ethtool} to modify the interface configuration should be examined. If the interface speed is toggling during a short period of time, the activity will be blocked. However, both user and kernel defensive components can be evaded by sophisticated malware such as rootkits. In addition, the malware may inject a shellcode with a signal generation code into a legitimate, trusted process to bypass the security products. To overcome the evasion techniques, it is possible to deploy solutions at the hypervisor level (Figure \ref{fig:filter}). In this approach, a hypervisor-level firewall inspects the changes to the network interface and examines the UDP packets from the active virtual machine. Irregular activities are logged, and the outgoing packets are blocked. Using visualization for firewalls and security studied in previous work \cite{brohi2012identifying}.

\subsection{Signal Monitoring} 
Another approach is to use RF monitoring hardware equipment to identify anomalies in the LANETNNA frequency bands. However, due to the legitimate activities of local network devices (e.g., UDP traffic) such a detection approach will lead to many false positives. 

\subsection{Signal Jamming} 
It is possible to block the covert channel by jamming the LANTENNA frequency bands. Modern jammers are signal blocking devices with radio frequency (RF) hardware that transmits radio waves in the entire range of the required frequency bands (e.g., the 250 MHz). A jammer generates high power, constant radio transmissions which span the channels and interrupt any covert channel transmissions. 
Another approach is to generate random traffic which interrupts the possible covert transmission from other devices in the network. In this case, a networked device such as a PC or Raspberry Pi generates UDP traffic at random times and in different volumes. 



\subsection {Cable Shielding} 
Metal shielding is a typical measure used to block or limit electromagnetic fields from interfering with or emanating from the shielded wires.  
Ethernet cable shielding copes with the threat presented in this paper by limiting
the leakage of signals generated by the LANTENNA techniques.
Different techniques can be used for shielding Ethernet cables. The most common is to place a shield around each twisted pair to reduce the general electromagnetic emission and the internal crosstalk between wires. It is possible to increase the protection by placing metal shielding around all the wires in the cable. Table \ref{tab:shields} contains the codes used to mark the different types of Ethernet cable shielding. 

\begin{table}[]
	\centering
	\caption{Ethernet cables shielding codes. TP = twisted pair, U = unshielded, F = foil shielded, S = braided shielding
	}
	\label{tab:shields}
	\begin{tabular}{@{}lll@{}}
		\toprule
		\# & Code  & Type of shielding                                   \\ \midrule
		1  & U/UTP & Unshielded cable, unshielded twisted pairs          \\
		2  & F/UTP & Foil shielded cable, unshielded twisted pairs       \\
		3  & U/FTP & Unshielded cable, foil shielded twisted pairs       \\
		4  & S/FTP &  Braided shielded cable, foil shielded twisted pairs \\ \bottomrule
	\end{tabular}
\end{table}


\section{Conclusion}
\label{sec:conclusion}
This paper shows that attackers can exploit the Ethernet cables to exfiltrate data from air-gapped networks. Malware installed in a secured workstation, laptop, or embedded device can invoke various network activities that generate electromagnetic emissions from Ethernet cables. We present two methods of signal generation: network speed toggling and UDP packet transmissions. We implemented malware (LANTENNA) and discussed the implementation details of the modulator and demodulator. We evaluated this covert channel in terms of bandwidth and distance and presented a set of countermeasures. Our results show that adversaries can transmit data several meters away from compromised air-gapped networks by using the electromagnetic covert channel. Furthermore, we show that this attack can be launched from an ordinary user-level process without root privileges and works successfully from within virtual machines.

\Urlmuskip=0mu plus 1mu\relax
\balance
\bibliographystyle{plain}
\bibliography{network,airfi,AirGap,AirGapCases,mobile,AirGapTools,PowerSupply,airgapsec,attackvectors}

\end{document}